# Molecular hyperdynamics coupled with the nonorthogonal tight-binding approach: Implementation and validation

K.P. Katin[1,2], K.S. Grishakov[1,2], A.I. Podlivaev[1,2], M.M. Maslov[1,2]

[1]*Department of Condensed Matter Physics, National Research Nuclear University 'MEPhI', Kashirskoe Shosse 31, Moscow 115409, Russia*
[2]*Laboratory of Computational Design of Nanostructures, Nanodevices and Nanotechnologies, Research Institute for the Development of Scientific and Educational Potential of Youth, Aviatorov Street 14/55, Moscow 119620, Russia*

ABSTRACT

We present the molecular hyperdynamics algorithm and its implementation to the nonorthogonal tight-binding model NTBM and the corresponding software. Due to its multiscale structure, the proposed approach provides the long time scale simulations (more than 1 s), unavailable for conventional molecular dynamics. No preliminary information about the system potential landscape is needed for the use of this technique. The optimal interatomic potential modification is automatically derived from the previous simulation steps. The average time between adjusted potential energy fluctuations provides an accurate evaluation of physical time during the hyperdynamics simulation. The main application of the presented hyperdynamics method is the study of thermal-induced defects arising in the middle-sized or relatively large atomic systems at low temperatures. To validate the presented method, we apply it to the $C_{60}$ cage and its derivative $C_{60}NH_2$. Hyperdynamics leads to the same results as a conventional molecular dynamics, but the former possesses much higher performance and accuracy due to the wider temperature region. The coefficient of acceleration achieves $10^7$ and more.

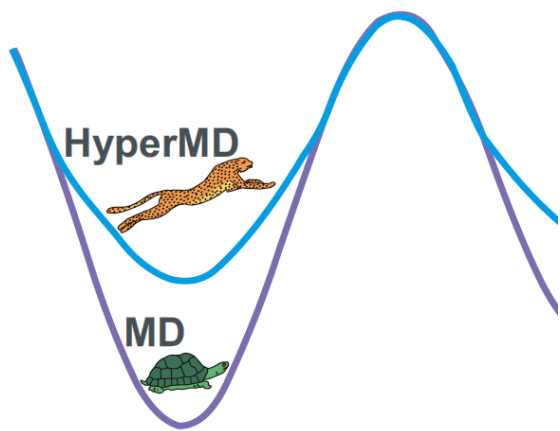

INTRODUCTION

Intelligibility of the molecular dynamics (MD) technique gives it a great advantage over the other methods of atomistic simulation. MD provides observation of the considered system evolution in a "real-time mode" and, therefore, it is the direct way to simulate any kinetic process occurring at the atomic level. In particular, MD is traditionally used for studying the thermal stability of nanostructures, nanomaterials, and biomolecules.[1-3]

On the other hand, the so-called "time-scale problem"[4,5] is a well-known drawback of traditional MD approach. This problem arises from a huge mismatch between typical MD time step ~0.1 fs, required for the adequate treating of atomic oscillations, and characteristic times of the studied processes, that can be rather macroscopic. Since the late 1990-s, different ways of traditional MD technique acceleration were developed. One of the series of the wildly used accelerated algorithms is based on the machine learning approach.[6,7] It provides the high-accuracy prediction of the molecular system potential landscape basing on the information derived from the previous MD steps. As a consequence, energy and atomic gradients are not recalculated at every step saving the computer resources.

Another basic idea related to the acceleration of MD treatment is the interatomic potential modification near the "deep wells", as illustrated at Fig.1. This trick dramatically reduces the time spending by the system to escape from the potential well. So, one can observe transitions from one state to another more frequently. Such an idea of potential modification produces two groups of techniques, known as hyperdynamics[4,8,9] and metadynamics[10,11]. In addition, their combinations are also presented.[12]

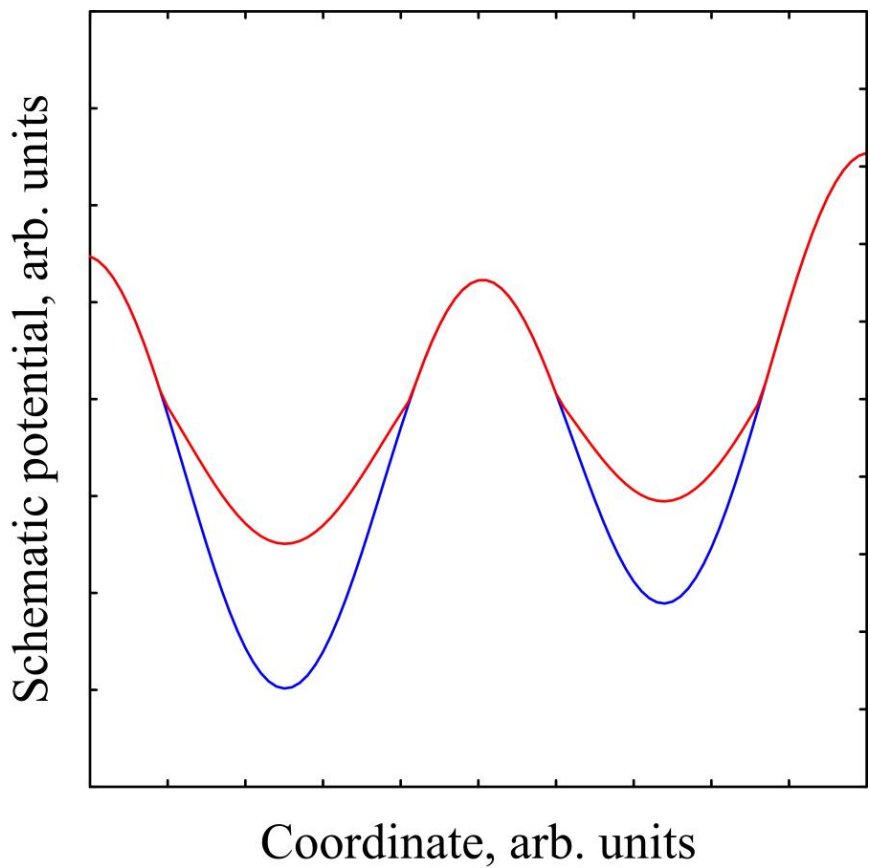

**Fig. 1**. Schematic illustration of pure (blue line) and modified (red line) interatomic potentials. Shallower wells provide more rapid system transitions from one state to another.

In the simplest form, hyperdynamics algorithm does not need any preliminary information about the system's potential landscape. This fact explains the application of hyperdynamics to many practically important problems, for example, water-driven degradation of polyamide[13], low-temperature decomposition of *n*-Heptane[14], crack propagation in elastic crystalline[15], and dissolution of organic crystals[16]. Metadynamics implies the introduction of the additional non-physical term into the system's Hamiltonian that forces the system to evolve in a particular direction (for example, along the assumed reaction coordinate). This method requires the obligatory preliminary investigation of the potential energy landscape and the collective variables and, therefore, is less favorable. Nevertheless, metadynamics was successfully applied to predict the rates of some physical, chemical, and biological processes, for example, the molecular condensation from the vapor[17], tryptophan-cage mini-protein transitions[18] and inhibitor activity of target p38 mitogen-activated protein kinase[19].

Both hyperdynamics and metadynamics approaches are based on the modified interatomic potential and are suitable for obtaining useful insight into the mechanisms of system possible transformations. Moreover, they are also applicable to computing the thermodynamic averages after the accurate reweighting of the accounted states whose energies were modified. However, it is difficult to compare the time step for the system evolved in modified potential, with the "real" time step in pristine potential. Another problem that is typical for these techniques is choosing the value of energy shift applied to the potential wells[20]: the very high shift can overcome the barriers between the different system states, whereas rather low shift can be inefficient.

Here we present some new kind of hyperdynamics as well as its implementation to the nonorthogonal tight-binding model NTBM and the corresponding software[21]. Our algorithm implies the automatic determination of the optimal potential modification. As it is common for the machine learning approaches, the information of the system's potential landscape is routinely accumulated during the simulation and then used in further steps. Multistep time rescaling provides the direct relationship between hyperdynamics time step and real physical time. The main advantage of the algorithm is the accurate evaluation of the period that is needed for the thermal-induced transformations occurring in the atomic system at low temperatures.

DESCRIPTION OF THE HYPERDYNAMICS ALGORITHM

The presented algorithm is designed for the simulation of atomic system evolution, including possible isomerization, defect formation, or decomposition. Due to the high performance of hyperdynamics, one can simulate all these processes at low temperatures. Introduction of wide temperature region leads to three advantages: (i) low temperatures are usually close to the real experimental conditions, (ii) low temperatures provide exclusion of the unfavorable defects and decomposition paths, arising exclusively at high temperatures, and (iii) wide temperature range results in more accurate kinetic parameters, which are derived from the temperature dependence of the reaction rate such as activation energy or frequency factor. To get the latter advantage, one should measure the physical time during the hyperdynamics simulation carefully. The described algorithm provides such ability via the comparison of the mean time between two adjacent fluctuations of the system potential energy.

At the initial step, $N$ traditional MD iterations are performed, and the maximum value $U_1$ of the system potential energy $U$ is defined. After that, traditional MD treatment continues until the system undergoes $M$ fluctuations of the potential energy (if $U > U_1$ during at least one hundred time steps, the event is accounted for as a fluctuation). Averaged time $t_1$ elapsed between two consequent fluctuations is calculated.

At the next step, the potential energy function is modified by the way typical for the hyperdynamics: if $U < U_1$, potential $U$ is replaced by the effective potential

$$U_{eff} = U_1 - \alpha(U_1 - U).$$

Coefficient $\alpha$ belongs to the range from zero to unity and tunes the flattering degree of the potential well: $\alpha = 1$ corresponds to the unmodified potential, whereas $\alpha = 0$ corresponds to the flat effective potential $U_{eff} = U_1$. In the modified potential, more sophisticated time accounting is realized: every time the system is located in the area of modified potential with $U < U_1$, time increases by $t_1$. Modified potential provides higher and more frequent fluctuations of potential energy. So, a new maximum of the potential energy $U_2$ during $N$ time steps in the non-modified region of phase space can be determined. After that, the mean time $t_2$ between two consequent fluctuations averaged over $M$ fluctuations, is calculated (if $U > U_2$ during at least one hundred time steps, the event is accounted for as a fluctuation). Next, the $U_2$ value is used for the modification of potential energy function instead of $U_1$: if $U < U_2$, potential $U$ is changed by

$$U_{eff} = U_2 - \alpha(U_2 - U).$$

The updated potential provides higher energy fluctuations, and one can derive another value set, $U_3$, and $t_3$, etc. In practice, up to ten of iterations are needed to achieve the desired event (defect, isomerization, or decomposition). Parameters $N$, $M$, and $\alpha$ determine the performance and convergence of the multistage algorithm.

IMPLEMENTATION TO THE NONORTHOGONAL TIGHT-BINDING MODEL

Hyperdynamics technique can be used along with any *ab initio*, semiempirical, or empirical interatomic potentials. To validate our hyperdynamics approach, we chose the nonorthogonal tight-binding model NTBM as a reasonable compromise between rigorous density functional theory and commonly used empirical force fields. The NTBM model was published in 2016,[21] but it was not implemented to any publically available software so far. Nevertheless, NTBM was successfully applied to several particular problems. It demonstrated very good agreement with *ab initio* results for the high-strained cage systems with unfavorable valence angles: CL-20 chains[22], cubanes[23], and hypercubane[24]. Here we present the NTBM software with the included hyperdynamics algorithm described above.

The NTBM package is available at https://www.ntbm.info. One can download it as a set of source code files or as a pre-compiled executable program for free. Source files are written using Fortran programming language. Common functional such as periodic boundary conditions, geometry optimization, frequencies calculation, and molecular dynamics, as well as hyperdynamics described above, are available. Additional functions will be progressively added.

## APPLICATION OF HYPERDYNAMICS TO THE $C_{60}$ AND $C_{60}NH_2$ CAGES

We apply hyperdynamics approach to simulate the evolution of two heated cage molecules: fullerene $C_{60}$ with $I_h$ symmetry and its derivative $C_{60}NH_2$. It is well known that the most feasible defect in $C_{60}$ is the so-called Stone-Wales transformation (rotation of one of the C–C bonds by ~90°  [25]). This transformation is schematically presented in Fig. 2a. Appearing of the Stone-Wales transformations leads to the distortion of carbon framework and finally results in the fullerene cage decomposition. For $C_{60}NH_2$, the most feasible process is the separation of $NH_2$ radical by the rapture of the C–N bond (see Fig. 2b). So, the thermal evolution of $C_{60}$ and $C_{60}NH_2$ cages introduce two common reaction types. They are the bond rotation and bond fission, respectively. That is why we choose these compounds to validate the hyperdynamics approach.

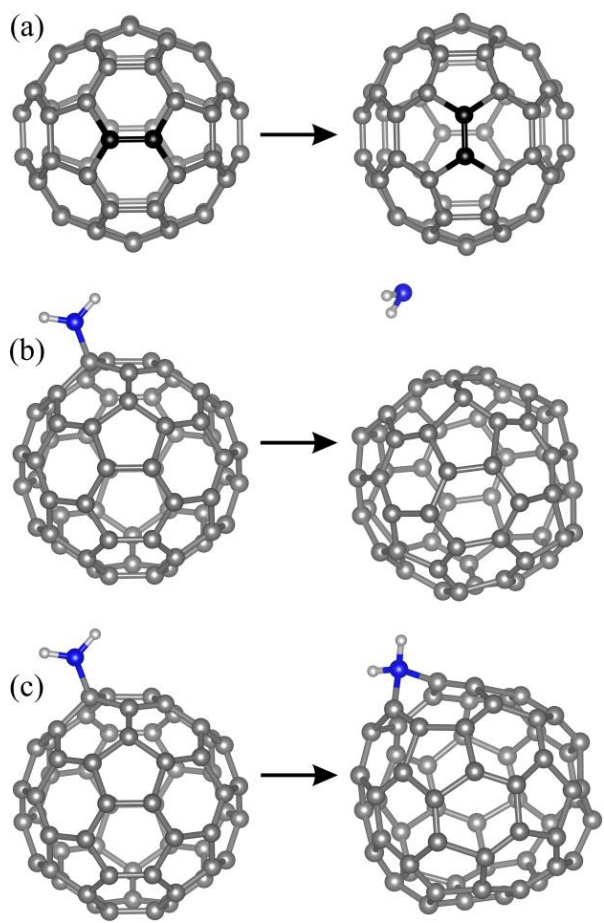

**Fig. 2**. The C–C bond rotation (Stone-Wales defect) in $C_{60}$ fullerene (a), the C–N bond fission in the functionalized $C_{60}NH_2$ cage (b), and bridge configuration with tetra-coordinated nitrogen that appeared at high temperatures (c).

Cages $C_{60}$ and $C_{60}NH_2$ were simulated using both classical MD and hyperdynamics approaches until the appearance of C–C bond rotation or C–N bond fission, respectively. Newton's equations of motion were integrated using the velocity Verlet algorithm with the time step of 0.1 fs. The constant temperature was maintained by the Andersen's thermostat[26], as it was implemented in NTBM software. Three independent runs with the different random velocity sets were performed at every temperature. Dimensionless parameters $N$, $M$, and $\alpha$ in hyperdynamics algorithm were equal to 10000, 100, and 0.5, respectively. For some set of temperatures, the additional calculations with $N = 20000$ and $M = 400$ were made, and the same reaction rates were obtained concerning the standard statistical deviations. Figure 3 presents the time that is needed for the considered reactions as a function of reverse temperature.

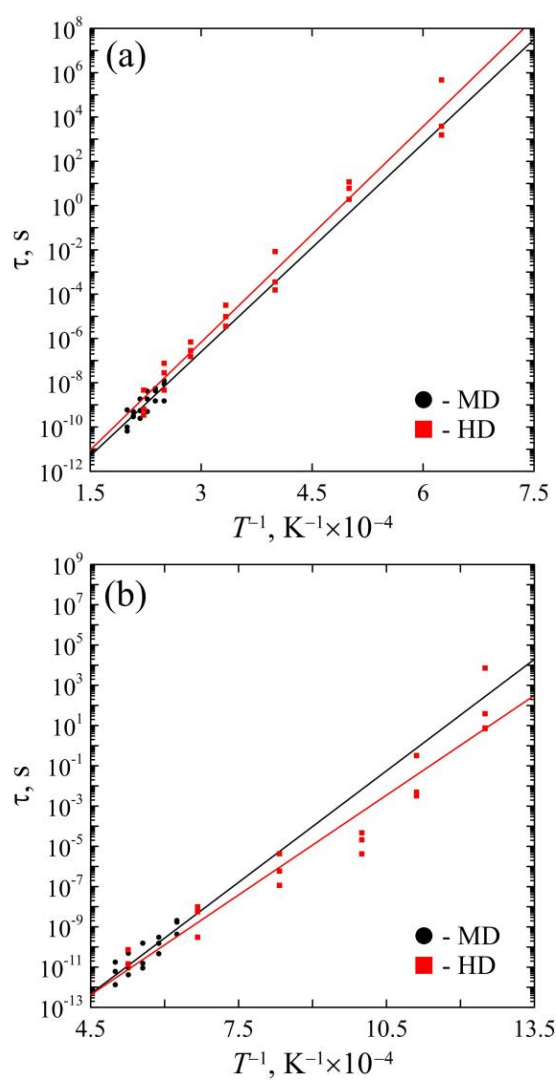

**Fig. 3**. The time required for the C–C bond rotation (Stone-Wales defect) in $C_{60}$ fullerene (a) and the C–N bond fission in the functionalized $C_{60}NH_2$ cage (b) as a function of reverse temperature $1/T$. Black circles and red squares are the calculation results in the frame of MD and hyperdynamics approaches, respectively. The lines are drawn through the points by the least-mean-squares method.

Straight lines are the linear approximations of these dependencies by the Arrhenius law

$$\tau^{-1} \sim \exp\left(-\frac{E_a}{k_B T}\right),$$

where $\tau$ is the time that is needed for the corresponding reaction, $E_a$ is the activation energy, $k_B$ is the Boltzmann's constant, $T$ is the temperature. Activation energies can be derived from the slopes of approximating lines. Both MD and hyperdynamics methods predict the formation of the same Stone-Wales defect imaged in Fig. 2a. We obtain the reasonably good agreement of reaction times predicted using these methods, see Fig. 2a. Corresponding activation energies are equal to 6.22 ± 0.99 and 6.44 ± 0.22 eV for MD and hyperdynamics, respectively, see Fig. 3a. They match with the previously reported corresponding energy barrier (6.48 eV [27]). One can see the high accuracy of hyperdynamics associated with the wide available temperature range. Note, that hyperdynamics provides reaction times of 1 s and more, whereas the comparable number of MD steps can provide only tens of nanoseconds. So, the acceleration factor due to the use of hyperdynamics reaches $10^7$ and more.

During the studying of heated $C_{60}NH_2$ cage, we observe two possible events: the C–N bond fission (Fig. 2b) and the formation of bridge configuration with tetra-coordinated nitrogen (Fig. 2c). The latter appears only at very high temperatures ($T \geq 1800$ K). Thus, we ignored this scenario in the activation energy evaluation. Such example confirms the importance of accounting for the low-temperature region to get the actual reaction mechanism. For the fission of the C–N bond, we obtain the activation energies of 3.66 ± 0.55 and 3.28 ± 0.21 eV for MD and hyperdynamics methods, respectively. These values are close to the previously reported binding energies between $NH_2$ functional group and small carbon nanotubes (2.5 ÷ 3.5 eV depending on the tube's chirality indices[28]). Figure 3b demonstrates reasonably well agreement between MD and hyperdynamics approaches.

CONCLUSION

Despite the progress in non-dynamical algorithms for the investigation of reactions paths, MD remains the most direct and reliable method to reveal the reaction mechanism or to probe the thermal stability for any atomic system. In many cases, accelerated MD techniques provide timescales comparable with the real experimental conditions. Concerning the extensive computational searching of advanced materials and compounds performed worldwide, accelerated MD can be widely applied to investigate their stability and possible thermal defect formation. Note that defects often determine the experimentally observed properties of real materials.

Here we presented the implementation of hyperdynamics algorithm coupled with the nonorthogonal tight-binding interatomic potential. The main advantages of this approach are (i) its applicability to any atomic system without the preliminary investigation of its potential landscape and (ii) direct relationship between every hyperdynamics step and real physical time. We believe that hyperdynamics can be useful in many problems of computational chemistry.

ACKNOWLEDGMENTS

The reported study was funded by RFBR according to the research project No. 18-32-20139 mol_a_ved. We are grateful to Maria Katina for the careful preparing of the graphical abstract.